\documentclass[10pt]{article}
\usepackage{OSAmeet}
\usepackage{amsbsy,latexsym}
\usepackage{amsfonts}
\usepackage{amssymb}
\usepackage[mathscr]{eucal}

\newcommand{\ket}[1]{\vert #1 \rangle}
\newcommand{\bra}[1]{\langle #1 \vert}

\begin{document}

\title{Communicating a direction using spin states}

\author{E.~Bagan, M.~Baig and R.~Mu{\~n}oz-Tapia}

\address{Grup de F{\'\i}sica Te{\`o}rica \& IFAE,\\ Facultat de
Ci{\`e}ncies, Edifici Cn, \\ Universitat Aut{\`o}noma de Barcelona,  \\
08193 Bellaterra (Barcelona) Spain}

\email{e-mail: bagan@ifae.es, baig@ifae.es, rmt@ifae.es}

\HeaderAuthorTitleMtg{Bagan et al.}{Communicating a direction
using spin states}{ICQI}

\begin{abstract}
The communication of directions using quantum states is a useful
laboratory test for some basic facts of quantum information. For a
system of spin-$1/2$ particles there are different quantum states
that can encode directions. This information can later be decoded
by means of a generalized measurement. In this talk we present
the optimal strategies  under different assumptions.
\end{abstract}


\noindent

Imagine two parties, traditionally called Alice and Bob. Alice
wants to communicate a space direction $\vec n$ to Bob, but she
only has at her disposal several spin-$1/2$ particles. She can
use them to construct a quantum state that will encode the
information about the direction. Upon receiving the state, Bob
performs a quantum measurement and tries to retrieve as much of
this information as possible. This simple scenario has a long
history and it has been invoked to test several hypothesis. The
first non trivial question is whether there are differences
between collective measurements as compared to repeated
individual measurements. In other words, is it better to perform
a global measurement on the quantum system as a whole or to
measure the spins separately? This was the original motivation of
Peres and Wooters~\cite{pw}. Later Massar and Popescu~\cite{mp}
using $N$ parallel spins concluded that collective measurements
are always better than individual ones.

A turn of events was the work of Gisin and Popescu~\cite{gp}.
These authors showed that, in the simple case of $N=2$, sending
the two spins in an antiparallel state leads to an even  greater
accuracy. This was the main motivation for our systematic study
of the optimal strategies for an arbitrary number of spins.

We start by describing the main elements of the scenario.
\begin{itemize}
\item {\it The encoding state}. Alice constructs a reference quantum
state $\ket{A}$ out of $N$ spin-$1/2$ particles. We take
$\ket{A}$  to be an eigenstate of $S_z$, i.e.,
$S_z\ket{A}=m\ket{A}$. Alice encodes the direction by performing a
unitary operation (a rotation) on the reference state,
$\ket{A}\mapsto\ket{A (\vec{n})}=U(\vec{n})\ket{A}$, where
$U(\vec{n})$ is generated by the spin operators $\vec{S}$. Hence,
one has $\vec{S}\cdot \vec{n}\ket{A(\vec{n})}=m\ket{A(\vec{n})}$.
\item {\it The measurements}. The most general measurement Bob can
perform is a positive operator valued measurement (POVM).
Mathematically, the POVM is defined by a set of positive
Hermitian operators $\{O_{r}\}$ that are a resolution of the
identity
                            $\mathbb I= \sum_{r} O_{r}$.
For each outcome $r$, Bob obtains a guess $\vec n_{r}$ for the
direction.
\item {\it The quality}. To quantify the quality of the guess, $\vec
n_{r}$, we use the fidelity  $f=(1+\vec{n}\cdot\vec{n}_r)/2$.
Thus, if Alice sends a number of isotropically distributed
directions, the average fidelity, that can be written as
\begin{equation}
F=\sum_r \int dn\frac{1+\vec{n}\cdot\vec{n}_r}{2}
\bra{A(\vec{n})} O_r\ket{A(\vec{n})}, \label{f1}
\end{equation}
 is a good figure of merit for the communicating
strategy ($dn$ is the rotationally invariant measure on the unit
two-sphere). It is known that  there exists an optimal continuous
POVM~\cite{holevo,us2} defined by a set of positive projectors of
the form $O(\vec n)=U(\vec n)\left[\ket{B}\bra{B}\right.+$\-
$\left.\ket{B'}\bra{B'}+\cdots\right]U^\dagger(\vec n)$, where
$U(\vec n)$ is the element of SU(2) associated with the rotation
$R:\vec z \mapsto \vec n$, and $\ket{B}$, $\ket{B'}$,~\dots, are
fixed states analogous to $\ket{A}$. Rotational invariance  then
enables us to write the average fidelity as
\begin{equation}
F=\int dn{1+\vec{z}\cdot\vec{n}\over 2} \bra{A}O(\vec{n})\ket{A}.
\label{fidelity}
\end{equation}
\end{itemize}

We now turn to the different strategies.

\noindent{\bf Parallel spins (P)} The first obvious approach
consists in sending identical copies of a single spin state. This
corresponds to taking Alice's reference state to be
$\ket{A}=\ket{\uparrow\uparrow \cdots \uparrow}$, where we use the
obvious notation $S_{z}\ket{\uparrow}=(1/2)\ket{\uparrow}$. Hence,
$\ket{A}=\ket{N/2,N/2}$ belongs to the highest spin
representation $\boldsymbol{J}=\boldsymbol{N/2}$. It is known
\cite{mp} that the maximal average fidelity (MAF) is
$F_P=(N+1)/(N+2)$ (this result follows directly from
Eq.~\ref{fidelity}), which is readily seen to approach one as
$F_P\sim 1-1/N$. Explicit theoretical realizations of the optimal
measurements with a finite number of outcomes were obtained
in~\cite{derka} for arbitrary $N$ and minimal versions of these
measurements for $N$ up to seven can be found in~\cite{lpt}.

\noindent{\bf Antiparallel spins (A)} Gisin and Popescu~\cite{gp}
showed that for $\ket{A}=\ket{\uparrow\downarrow}$ the MAF reads
$F=(3+\sqrt{3})/6 > 3/4$. In \cite{massar,us2} it is proved that,
indeed, this is the optimal result for $N$=2. How to proceed
beyond $N=2$? A straightforward generalization would be to take
$\ket{A}=\ket{\uparrow\downarrow\downarrow\cdots\uparrow}$, with
$n_{\uparrow}$ spins up and $n_{\downarrow}$ spins down (we will
loosely refer to them as product states). These states belong to
the direct sum of irreducible representations given by the
Clebsch-Gordan decomposition $\boldsymbol{(1/2)}^{\otimes N}=
\boldsymbol{N/2\oplus (N/2-1)\oplus \cdots}$, where equivalent
representations occur more than once  except for {\boldmath
$N/2$}. Similarly, the encoding rotations are written as $U(\vec
n)=\bigoplus_{j} U^{(j)}(\vec n)$. The MAF in (\ref{fidelity})
can be computed using the effective state $\ket{\tilde
A}=\sum_{j=m}^{N/2}\tilde{A}_j \ket{j,m}$, where
$m=(n_{\uparrow}-n_{\downarrow})/2$, and the coefficients
$\tilde{A}_j$ are explicitly given by~\cite{us3}
\begin{equation}\label{Aj-2}
 \tilde{A}_j=\sqrt{\frac{1+2j}{J+1+j}}\sqrt{\frac{(J-m)!(J+m)!}{(J-j)!(J+j)!}}
;\quad J\equiv {N\over2}.
\end{equation}
It is important to notice that only one of each equivalent
irreducible representation appears in $\ket{\tilde A}$. Similarly,
for the POVM it is sufficient to use just a single effective state
$\ket{\tilde B}=\sum_{j=m}^J  \sqrt{2j+1}\ket{j,m}$, with
$J=N/2$. In \cite{us3} we showed that the MAF is attained for the
minimal $m$ ($m=0$ for $N=2n$, and $m=1/2$ for $N=2n+1$), i.e.,
for states with $n_{\uparrow}$ as close as possible to
$n_{\downarrow}$, as hinted from the $N=2$ case. For $N=2n$ the
MAF (\ref{fidelity}) takes the simple form
\begin{equation}\label{fidelity-3}
F_{A}=\frac{1}{2}+\sum_{j=1}^{n} \frac{n!^2}{(n-j)! (n+j)!}
\frac{j}{\sqrt{(n+1)^2-j^2}}.
\end{equation}
We have collected the results of $F_A$ for $N\leq 7$ in Table~1.
One can show  that $F_A>F_P$ for any $N$, and that $F_A$
approaches unity faster that $F_P$, specifically one has $F_A \sim
1-1/(2N)$.

\noindent{\bf Optimal states (O)} An obvious improvement on the
previous strategy is obtained if one relaxes the condition that
$\ket{A}$ is a product state. In other words, we  only require
$\ket{A}$ be a normalized state and we let the maximization
conditions to fix its components. The optimal POVM is the same as
in the previous case and the MAF is also obtained for the minimal
values of $m$, ($m=0,1/2$ for an even/odd number of spins).
However the optimal states $\ket{A}$ are in general {\em
entangled}. The results of the MAF, which we denote here with the
suffix $O$ (optimal), can be written as~\cite{us2}
\begin{equation}\label{fo}
F_{O} =  \frac{1}{2}(1+x^{0,0}_{N/2+1}),\quad  \mbox{for $N$
even};\qquad F_{O} = \frac{1}{2}(1+x^{0,1}_{N/2+1/2}),\quad
\mbox{for $N$ odd},
\end{equation}
where~$x^{0,0}_{N/2+1}$ ($x^{0,1}_{N/2+1/2}$) is the largest zero
of the Legendre polynomial $P_{N/2+1}=P^{0,0}_{N/2+1}$  (Jacobi
polynomial $P_{N/2+1/2}^{0,1}$). A numerical analysis can be found
in~\cite{ps}. Note that $F_O$ is always greater than $ F_A$
(except for $N=2$ where both coincide). The asymptotic behaviour
of $F_O$ is also qualitatively different: it approaches unity
quadratically in the number of spins, $ F\sim 1-\xi^2/N^2$, where
$\xi\sim 2.4$ is the first zero of the Bessel function $J_0$
(recall that  for product states the fidelity approaches unity
only linearly). {\em Actually, it is precisely this different
asymptotic behaviour what proves that the optimal states must be
entangled}. Furthermore, the asymptotic behaviour can be
understood in terms of the dimension $d$ of the Hilbert space
{\em effectively} used in each case. Encoding with $N$ parallel
spins uses only  the representation~{\boldmath $N/2$}, whose
dimension is $d=N+1$ (for the antiparallel case one also has
$d\sim N$), whereas the optimal strategy uses a much larger
Hilbert space, with $d\sim N^2$.

\noindent{\bf General encodings (G)} The strategies presented so
far have the property that the direction is encoded {\em in an
intrinsic way}, as the states $\ket{A(\vec{n})}$ are the quantum
analog of the classical gyroscopes. The unitary operations
$U(\vec{n})$ are group representations of spatial rotations.
Hence, for instance, Alice may encode the direction by physically
rotating the device that she uses to prepare her reference state
$\ket{A}$. In this sense, (\ref{fo}) is the maximal fidelity that
can possibly be achieved by any intrinsic strategy, i.e., those
that do not require the existence of a common reference frame
shared by Alice and Bob. Nevertheless, since the dimension is the
responsible for the improvement in the fidelity, it is natural to
search also for more general strategies that use up the Hilbert
space dimension of the encoding states~\cite{us1}. For instance,
for a system of two spin-$1/2$ particles one has $d=4$. It is
possible to construct in this Hilbert space generators $\vec{S}$
that belong to the $\boldsymbol{3/2}$ (4-dimensional)
representation of SU(2), as if we actually had a single
spin-$3/2$ particle~\cite{Pe, us2}, and despite of the fact that
$\boldsymbol{1/2\otimes1/2}=\boldsymbol{1\oplus0\not}=\boldsymbol{3/2}$.
Furthermore $\vec{S}$ {\em can not be the total spin operators of
the two-particle state}, which, according to the Clebsch-Gordan
decomposition above, should belong to the {\boldmath $1 \oplus 0$}
representation. Thus, $U(\vec{n})$ is not a representation of a
spatial rotation, and $\vec{S}$ is not a real vector, therefore
Alice and Bob must share a reference frame to specify $\vec{S}$.
The expression of the MAF is formally equal to that of the
parallel encoding with $N=3$, i.e, $F^{(d=4)}=4/5$. This result
can be generalized to an arbitrary dimension: the single
spin-$(d-1)/2$ interpretation of a $d$-dimensional Hilbert space
gives the optimal encoding with a MAF $F^{(d)}={d/(d+1)}$. If
$d=2^{N}$, one can, of course, perform this  encoding with $N$
spin-$1/2$ particles obtaining a fidelity $F_{G}=2^N/(2^N+1)$.
The behaviour of the fidelity as a function of the number of spins
tends to unity exponentially: $F\sim 1-2^{-N}$. In terms of the
fidelity attained, this approach, although not intrinsic, is the
true optimal one.
\begin{table}\label{table-I}
\begin{center}
\begin{tabular}{c|ccccccl}
  \hline
  \hline
  $N$  & 2 & 3 & 4 & 5 & 6 & 7 & Large $N$ \\
  \hline
$F_P$ & 0.75   & 0.8    & 0.8333 & 0.8571 & 0.875  & 0.8889 &
$1-1/N$
\\
$F_A$ & 0.7887 & 0.8444 & 0.8848 & 0.9069 & 0.9235 & 0.9342 &
$1-1/(2N)$
\\
$F_O$ & 0.7887 & 0.8449 & 0.8873 & 0.9114 & 0.9306 & 0.9429 &
$1-\xi^2/N^2$
\\
$F_G$ & 0.8    & 0.8889 & 0.9412 & 0.9697 & 0.9846 & 0.9922 &
$1-1/2^N$
\\
  \hline
  \hline
\end{tabular}
\end{center}
\caption{Maximal average fidelities ($F$) for parallel ($P$),
antiparallel ($A$), optimal antiparallel ($O$) and general ($G$)
encodings. Parameter $\xi\sim 2.4$ is defined in the text.}
\end{table}

In summary, we have presented four different ways of
communicating a direction using a quantum channel based on $N$
spin-$1/2$ particles. We paid special attention to the large $N$
behaviour of the corresponding fidelities  (see Table). By
comparing $F_{A}$ and $F_{O}$ one readily sees that the optimal
encoding O requires entangled states. Though this has been argued
to be the case, to the best of our knowledge, we here provide the
first real proof of this statement.

We thank R. Tarrach and A. Brey for their collaboration in several
stages of this work. Financial support from CICYT contract
AEN99-0766 and CIRIT contracts 1998SGR-00051, 2000SGR-00063 is
acknowledged.

\end{document}